\documentclass[11pt]{article}%
\usepackage{amsfonts}
\usepackage{amsmath}
\usepackage{amssymb}
\usepackage{graphicx}
\usepackage{hyperref}%
\providecommand{\U}[1]{\protect\rule{.1in}{.1in}}
\begin{document}

\title{Cayley transforms of $su(2)$ representations}
\author{{\Large T. S. Van Kortryk\medskip}\\120 Payne Street, Paris MO65275\\{\small vankortryk@gmail.com}}
\date{}
\maketitle

\begin{abstract}
Cayley rational forms for rotations are given as explicit matrix polynomials
for any quantized spin $j$. \ The results are compared to the
Curtright-Fairlie-Zachos matrix polynomials for rotations represented as exponentials.

\end{abstract}

Curtright, Fairlie, and Zachos (CFZ) obtained an elegant and intuitive result
\cite{CFZ} expressing the exponential form for a rotation about an
axis\ $\boldsymbol{\hat{n}}$, for \emph{any} quantized angular momentum $j$,
as a polynomial of order $2j$ in the corresponding $\left(  2j+1\right)
\times\left(  2j+1\right)  $ spin matrices $\boldsymbol{\hat{n}\cdot J}$,
thereby culminating several decades of previous studies on this or closely
related problems \cite{Wigner}-\cite{T}. \ For each angle-dependent
coefficient of the polynomial, the explicit formula found by CFZ involves
nothing more complicated than a truncated series expansion for a power of the
$\arcsin$ function. \ Although a detailed proof of the CFZ result is not
exhibited in \cite{CFZ}, the essential ingredients needed to provide such a
proof are in that paper, and indeed, details of two elementary derivations
were subsequently given in \cite{CvK}. \ 

I discuss here the corresponding polynomial result for the Cayley rational
form of an irreducible, unitary SU(2) representation with spin $j$. \ In this
case, the explicit polynomial coefficients involve nothing more complicated
than truncated series expansions for simple, finite products. \ I compare the
Cayley form to the CFZ result.

The CFZ formula for a rotation through an angle $\theta$ about an axis
$\boldsymbol{\hat{n}}$, valid for any spin $j\in\left\{  0,\frac{1}{2}%
,1,\frac{3}{2},\cdots\right\}  $, is given by the unitary $SU\left(  2\right)
$ matrix%
\begin{equation}
U=\exp\left(  i~\theta~\boldsymbol{\hat{n}\cdot J}\right)  =\sum_{k=0}%
^{2j}\frac{1}{k!}\left.  A_{k}^{\left[  j\right]  }\left(  \theta\right)
\right.  \left(  2i~\boldsymbol{\hat{n}\cdot J}\right)  ^{k}~,
\label{CFZ result}%
\end{equation}
where the angle-dependent coefficients of the various spin matrix powers are
explicitly given by%
\begin{equation}
A_{k}^{\left[  j\right]  }\left(  \theta\right)  =\sin^{k}\left(
\theta/2\right)  ~\left(  \cos\left(  \theta/2\right)  \right)  ^{\epsilon
\left(  j,k\right)  }~\operatorname*{Trunc}_{\left\lfloor j-k/2\right\rfloor
}\left[  \frac{1}{(\sqrt{1-x})^{\epsilon\left(  j,k\right)  }}\left(
\frac{\arcsin\sqrt{x}}{\sqrt{x}}\right)  ^{k}\right]  _{x=\sin^{2}\left(
\theta/2\right)  }\ . \label{CFZ coefficients}%
\end{equation}
Here, $\left\lfloor \cdots\right\rfloor $ is the integer-valued floor function
while $\epsilon\left(  j,k\right)  $ is a binary-valued function of $2j-k$
that distinguishes even and odd integers: $\ \epsilon\left(  j,k\right)  =0$
for even $2j-k$, and $\epsilon\left(  j,k\right)  =1$ for odd $2j-k$. \ More
importantly, $\operatorname*{Trunc}\limits_{n}\left[  f\left(  x\right)
\right]  $ is the $n$th-order Taylor polynomial truncation for any $f\left(
x\right)  $ admitting a power series representation:%
\begin{equation}
f\left(  x\right)  =\sum_{m=0}^{\infty}f_{m}x^{m}%
\ ,\ \ \ \operatorname*{Trunc}_{n}\left[  f\left(  x\right)  \right]
\equiv\sum_{m=0}^{n}f_{m}x^{m}\ .
\end{equation}

To emulate the CFZ formula, I now construct
\href{http://en.wikipedia.org/wiki/Cayley_transform}{Cayley transforms}
\cite{Cayley,CayleyTransform} of elements in the $su\left(  2\right)
$\ algebra as spin matrix polynomials.

At first glance the Cayley transform for any spin representation would seem to
follow immediately from the CFZ result just by changing variables from
$\theta$ to a function of the angle, $\alpha\left(  \theta\right)  $. \ For
any given numerical value of $\boldsymbol{\hat{n}\cdot J}$\ this would be so,
of course, but it is not obviously so for matrix-valued $\boldsymbol{\hat
{n}\cdot J}$. \ In fact, it turns out that the matrix produced by the Cayley
transform of an $su\left(  2\right)  $ generator reduces to a spin matrix
polynomial that has a simpler form than the CFZ result (\ref{CFZ result}) for
the exponential of that generator.

For a finite dimensional, unitary, irreducible representation of $SU\left(
2\right)  $, all the underlying spin matrices $iS\equiv2i~\boldsymbol{\hat
{n}\cdot J}$ are anti-hermitian, and for spin $j$ satisfy the
\href{http://en.wikipedia.org/wiki/Cayley-Hamilton_theorem}{Cayley-Hamilton
theorem} \cite{Cayley,CayleyHamilton}\ appropriate for $\left(  2j+1\right)
\times\left(  2j+1\right)  $ matrices. \ Consequently the Cayley rational form
of a unitary $SU\left(  2\right)  $ group element for spin $j$ can be reduced
to a spin matrix polynomial \cite{NB}:%
\begin{equation}
U=\frac{1+2i\alpha~\boldsymbol{\hat{n}\cdot J}}{1-2i\alpha~\boldsymbol{\hat
{n}\cdot J}}=\sum_{k=0}^{2j}\left.  \mathfrak{A}_{k}^{\left[  j\right]
}\left(  \alpha\right)  \right.  \left(  2i~\boldsymbol{\hat{n}\cdot
J}\right)  ^{k}\ , \label{Cayley form}%
\end{equation}
where $\alpha$ is a real parameter, and where the coefficients $\mathfrak{A}%
_{k}^{\left[  j\right]  }\left(  \alpha\right)  $\ are to be determined as
functions of $\alpha$. \ The challenge here is to rewrite the geometric series
$1/\left(  1-2i\alpha~\boldsymbol{\hat{n}\cdot J}\right)  $ for spin $j$ as a
polynomial in $\boldsymbol{\hat{n}\cdot J}$. \ Thus define%
\begin{equation}
\frac{1}{1-2i\alpha~\boldsymbol{\hat{n}\cdot J}}=\sum_{k=0}^{2j}\left.
\mathfrak{B}_{k}^{\left[  j\right]  }\left(  \alpha\right)  \right.  \left(
2i~\boldsymbol{\hat{n}\cdot J}\right)  ^{k}\ .
\label{GeometricSeriesPolynomial}%
\end{equation}
Then clearly $\mathfrak{A}_{0}^{\left[  j\right]  }=2\mathfrak{B}_{0}^{\left[
j\right]  }-1$ and $\mathfrak{A}_{k\geq1}^{\left[  j\right]  }=2\mathfrak{B}%
_{k\geq1}^{\left[  j\right]  }$. \ The coefficients in the latter expansion
follow directly from the methods in \cite{CFZ,CvK}. \ The results are
succinctly given by%
\begin{equation}
\mathfrak{B}_{k}^{\left[  j\right]  }\left(  \alpha\right)  =\frac{\alpha^{k}%
}{\det\left(  1-2i\alpha~\boldsymbol{\hat{n}\cdot J}\right)  }%
~\operatorname*{Trunc}_{2j-k}\left[  \det\left(  1-2i\alpha~\boldsymbol{\hat
{n}\cdot J}\right)  \right]  \ , \label{GSPCoefficients}%
\end{equation}
where the truncation is in powers of $\alpha$, and where the determinant for
spin $j$ is%
\begin{equation}
\det\left(  1-2i\alpha~\boldsymbol{\hat{n}\cdot J}\right)  =%
{\displaystyle\prod\limits_{m=0}^{2j}}
\left(  1-2i\alpha\left(  j-m\right)  \right)  =%
{\displaystyle\prod\limits_{n=1}^{\left\lfloor j+1/2\right\rfloor }}
\left(  1+4\alpha^{2}\left(  j+1-n\right)  ^{2}\right)  \ .
\label{Determinant}%
\end{equation}
These results are readily checked for small values of $j$ upon using explicit
matrices, say $\boldsymbol{\hat{n}\cdot J}=J_{3}$. \ Indeed, this is how
(\ref{GSPCoefficients}) was first deduced. \ 

Subsequently, after the first version of this paper was posted on the arXiv, a
detailed proof\ of (\ref{GSPCoefficients}) was given in \cite{TLC}. \ But, as
it turns out, after the previous version of this paper was posted, I learned
that (\ref{GeometricSeriesPolynomial}) and (\ref{GSPCoefficients})\ are true
for any $N\times N$ matrix $M$, upon replacing $\boldsymbol{\hat{n}\cdot
J}\rightarrow M$ and $2j+1\rightarrow N$. \ The result is a widely known
theorem from the resolvent formalism (e.g. see \cite{He}), and is proven in
numerous textbooks and articles on matrix theory (e.g. see \cite{Householder}
and \cite{Hou}).

Still, some comments are in order. \ Firstly, note that for either bosonic
(integer) or fermionic (semi-integer) spins, only \emph{even} powers of
$\alpha$ with positive coefficients are produced by the determinant factors in
(\ref{GSPCoefficients}). \ Consequently the $\mathfrak{A}_{k}^{\left[
j\right]  }\left(  \alpha\right)  $ and $\mathfrak{B}_{k}^{\left[  j\right]
}\left(  \alpha\right)  $\ coefficients have no singularities for real
$\alpha$. \ Secondly, the determinants in (\ref{GSPCoefficients}) are
essentially generating functions of the central factorial numbers $t\left(
m,n\right)  $ (see \cite{CFN}), a fact already exploited in \cite{CFZ,CvK}.
\ For example, for integer $j$,%
\begin{equation}
\operatorname*{Trunc}_{2j-k}\left[  \det\left(  1-2i\alpha~\boldsymbol{\hat
{n}\cdot J}\right)  \right]  =\sum_{m=0}^{\left\lfloor j-k/2\right\rfloor
}4^{m}\alpha^{2m}~\left\vert t\left(  2j+2,2j+2-2m\right)  \right\vert \ ,
\label{asCFNs}%
\end{equation}
with the full determinant obtained for $k=0$. \ Thirdly, as $j\rightarrow
\infty$ for any fixed $k$ the truncation in (\ref{GSPCoefficients}) is lifted
--- albeit not without some subtleties \cite{LargeSpin} ---\ to obtain, for
small $\alpha$, $\lim_{j\rightarrow\infty}\mathfrak{B}_{k}^{\left[  j\right]
}\left(  \alpha\right)  \sim\alpha^{k}$. \ Nevertheless, in contrast to the
periodicized $\theta$-monomials found in \cite{CFZ}, the large $j$ behavior
here does not make the periodicity of rotations manifest. \ To exhibit
periodicity even for finite values of $j$, $\theta$ must be expressed, on a
case-by-case basis, as cyclometric functions of $\alpha$, and then the branch
structure of those cyclometric functions must be invoked.

Finally, the
\href{https://en.wikipedia.org/wiki/Hille-Yosida_theorem}{Hille-Yosida
theorem} from
\href{https://en.wikipedia.org/wiki/Resolvent_formalism}{resolvent theory}
provides a one-to-one relation between (\ref{CFZ coefficients}) and
(\ref{GSPCoefficients}). \ That is, for any hermitian $N\times N$ matrix, a
Laplace transform gives%
\begin{equation}
\frac{1}{1-itM}=\int_{0}^{\infty}e^{-s}e^{istM}ds\ .\label{HY}%
\end{equation}
So for linearly independent powers $M^{k}$, $0\leq k<N$, as is the case for
spin matrices, the matrix polynomial expansion coefficients are related
order-by-order.%
\begin{equation}
\frac{1}{1-itM}=\sum_{n=0}^{N-1}A_{n}\left(  t\right)  \left(  iM\right)
^{n}\ ,\ \ \ \exp\left(  i\phi M\right)  =\sum_{n=0}^{N-1}C_{n}\left(
\phi\right)  \left(  iM\right)  ^{n}\ ,\ \ \ A_{n}\left(  t\right)  =\int%
_{0}^{\infty}e^{-s}C_{n}\left(  st\right)  ds\ .
\end{equation}
This provides yet another route to prove the CFZ result (\ref{CFZ result})
starting from the known coefficients for the resolvent\ (see \cite{TLC}).
\ For example, for spin $j=1/2$,%
\begin{gather}
\exp\left(  \tfrac{1}{2}~i\phi~\sigma_{3}\right)  =\cos\left(  \tfrac{\phi}%
{2}\right)  ~I+i\sin\left(  \tfrac{\phi}{2}\right)  ~\sigma_{3}\ ,\\
\nonumber\\
\int_{0}^{\infty}e^{-s}\cos\left(  st/2\right)  ds=\frac{1}{1+\frac{1}%
{4}~t^{2}}\ ,\ \ \ \int_{0}^{\infty}e^{-s}\sin\left(  st/2\right)
ds=\frac{\frac{1}{2}~t}{1+\frac{1}{4}~t^{2}}\ ,\\
\nonumber\\
\det\left(  1-\tfrac{1}{2}~it~\sigma_{3}\right)  =1+\tfrac{1}{4}~t^{2}\ ,\\
\nonumber\\
\frac{1}{1-\tfrac{1}{2}~it~\sigma_{3}}=\frac{1}{1+\frac{1}{4}~t^{2}}\left(
I+\tfrac{1}{2}~it~\sigma_{3}\right)  \ ,
\end{gather}
where $I=\left(
\begin{array}
[c]{cc}%
1 & 0\\
0 & 1
\end{array}
\right)  $ and the Pauli matrix is $\sigma_{3}=\left(
\begin{array}
[c]{cc}%
1 & 0\\
0 & -1
\end{array}
\right)  $, as usual.

\paragraph*{Acknowledgements}

I thank Curtright and Zachos for suggesting this problem, and for discussions
and encouragement. \ I also thank Jack and Peggy Nichols for their
hospitality. \ Finally, I have benefited from stimulating views of sea waves
approaching the Malec\'{o}n, La Habana.\bigskip

\begin{center}

\end{center}


\begin{thebibliography}{99}
\bibitem {LargeSpin}The subtleties involved in the limit of large $j$ are
discussed more fully in \cite{TLC} where the precise meaning of $\lim
_{j\rightarrow\infty}\mathfrak{B}_{k}^{\left[  j\right]  }\left(
\alpha\right)  \sim\alpha^{k}$\ is explained in detail, and where exact
$\lim_{j\rightarrow\infty}\mathfrak{B}_{k}^{\left[  j\right]  }\left(
\alpha\right)  $ results are obtained for both integer and semi-integer $j$.
\end{thebibliography}
\end{document}